\documentclass[%
 reprint,
 amsmath,amssymb,
 aps,
]{revtex4-2}

\usepackage{graphicx}
\usepackage{dcolumn}
\usepackage{bm}
\usepackage[percent]{overpic}
\usepackage{units}

\newcommand{\pt}{p_\mathrm{T}}
\newcommand{\mt}{m_\mathrm{T}}
\newcommand{\betat}{\beta_\mathrm{T}}
\newcommand{\rhat}{\hat r}
\newcommand{\phihat}{\hat \phi}
\newcommand{\phib}{\phi_\mathrm{b}}
\newcommand{\phip}{\phi_\mathrm{p}}
\newcommand{\phis}{\phi_\mathrm{s}}
\newcommand{\xip}{\xi_\mathrm{p}}
\newcommand{\xim}{\xi_\mathrm{m}}
\newcommand{\rmd}{\mathrm{d}}

\newcommand{\tsc}[1]{\textsc{#1}}
\newcommand{\Py}{\tsc{Pythia}}

\DeclareMathOperator{\arctanh}{arctanh}

\begin{document}



\title{Blast-wave description of $\Upsilon$ elliptic flow at energies available at the CERN Large Hadron Collider
}

\author{Klaus Reygers}
\author{Alexander Schmah}%
\affiliation{%
 Physikalisches Institut, Ruprecht-Karls-Universit\"at Heidelberg, Heidelberg, Germany
}%

\author{Anastasia Berdnikova}
\affiliation{
 National Research Nuclear University MEPhI, Moscow, Russian Federation
}%

\author{Xu Sun}
\affiliation{%
 Georgia State University, Atlanta, Georgia 30303, USA
}%

\date{\today}

\begin{abstract}
A simultaneous blast-wave fit to particle yields and elliptic flow ($v_{2}$) measured as a function of transverse momentum in Pb--Pb collisions at energies available at the Large Hadron Collider (LHC) is presented. A compact formula for the calculation of $v_2(\pt)$ for an elliptic freeze-out surface is used which follows from the Cooper-Frye ansatz without further assumptions. Over the full available $\pt$ range, the $\Upsilon$ elliptic flow data is described by the prediction based on the fit to lighter particles. This prediction shows that, due to the large $\Upsilon$ mass, a sizable elliptic flow is only expected at transverse momenta above 10\,GeV/$c$.
\end{abstract}

\maketitle


\section{\label{sec:introduction} Introduction}
Hydrodynamic models describing the space-time evolution of the medium created in ultra-relativistic nucleus-nucleus collisions explain many features of the observed particle spectra \cite{Huovinen:2006jp,Heinz:2009xj,Gale:2013da}. The blast wave model \cite{Schnedermann:1992ra,Schnedermann:1993ws} is a simple and computationally inexpensive model capable of capturing important aspects of hydrodynamic models. It is frequently used to extract information on the collective radial velocities and the decoupling temperature of the medium from measured transverse momentum spectra \cite{Abelev:2008ab,Abelev:2013vea,Khachatryan:2016yru,Acharya:2019yoi}. It provides a good description of $\pt$ spectra for center-of-mass energies ranging from a few GeV up to LHC energies. Blast-wave models were also used in the past to fit elliptic flow data for various collision systems and energies~\cite{Adler:2001nb,He:2010vw,Sun:2014rda}.

For pions, kaons, and protons, the blast-wave model provides a good description of the spectra for low transverse momenta ($\pt \lesssim \unit[2]{GeV}/c$). These particles are assumed to originate from a thermalized source in this $\pt$ range and the contribution from fragmenting quarks and gluons from hard scattering processes are expected to be small. The blast-wave model was also used to describe transverse momentum spectra of heavier particles like $J/\psi$ mesons and $\Omega$ baryons, where it is often assumed that different particle species freeze-out from the fireball at different temperatures \cite{Bugaev:2002fd,He:2010vw,Bellwied:2013cta,Andronic:2019wva}.

Measurements of spectra and elliptic flow of $J/\psi$ and $\Upsilon$ mesons are important for addressing the question to what extent also the heavy $c$ and $b$ quarks thermalize in the quark-gluon plasma and take part in the collective expansion of the fireball \cite{Das:2018xel}. $J/\psi$ data are reproduced by models which assume a large fraction of $J/\psi$ mesons to be regenerated from charm quarks in the plasma. This happens either within the plasma evolution \cite{Du:2015wha,Zhou:2014kka} or at the transition to the hadron gas \cite{Andronic:2019wva}. Owing to their large mass, $b$ quarks might not necessarily thermalize and consequently the regeneration component of $\Upsilon$ mesons is expected to be smaller than for $J/\psi$ mesons \cite{Du:2017qkv}. However, the $\Upsilon(2S)/\Upsilon(1S)$ ratio in Pb--Pb collisions at $\sqrt{s_\mathrm{NN}} = \unit[2.76]{TeV}$ can be described in a statistical model approach \cite{Andronic:2017pug}, an observation which would naturally go with a collective behavior of $\Upsilon$ mesons. The ALICE Collaboration recently published $\Upsilon$ elliptic flow data in Pb--Pb collisions at 5.02 TeV~\cite{Acharya:2019hlv}. The $v_2$ was found to be consistent with zero over the measured $\pt$ range. The $\Upsilon$ is therefore the first measured hadron species that does not exhibit elliptic flow in Pb--Pb collisions at the LHC. The data were found to be in agreement with transport calculations in which the fraction of regenerated $\Upsilon$ mesons is small \cite{Du:2017qkv,Bhaduri:2018iwr}.

In this paper, we describe a simultaneous blast wave fit to transverse momentum spectra and $v_2(\pt)$ including light ($\pi$, K, p) and heavy particles. This serves as a baseline model which helps to identify to what extent heavier particles behave differently from ($\pi$, K, p). We focus in particular on $\Upsilon$ mesons. Under the assumption that $b$ quarks fully thermalize, one would expect that also the $\Upsilon$ spectra and $v_2$ can be described by the blast wave model. By confronting the blast wave model predictions with data, we study whether this extreme assumption can be ruled out by currently available data. In order to perform the simultaneous blast wave fit to spectra and $v_2$, we derive a compact and easy-to-evaluate formula for $v_2(\pt)$.


\section{\label{sec:formula} Blast-wave model with elliptical freeze-out hyper-surface}

\begin{figure}
\includegraphics[width = 0.5\linewidth]{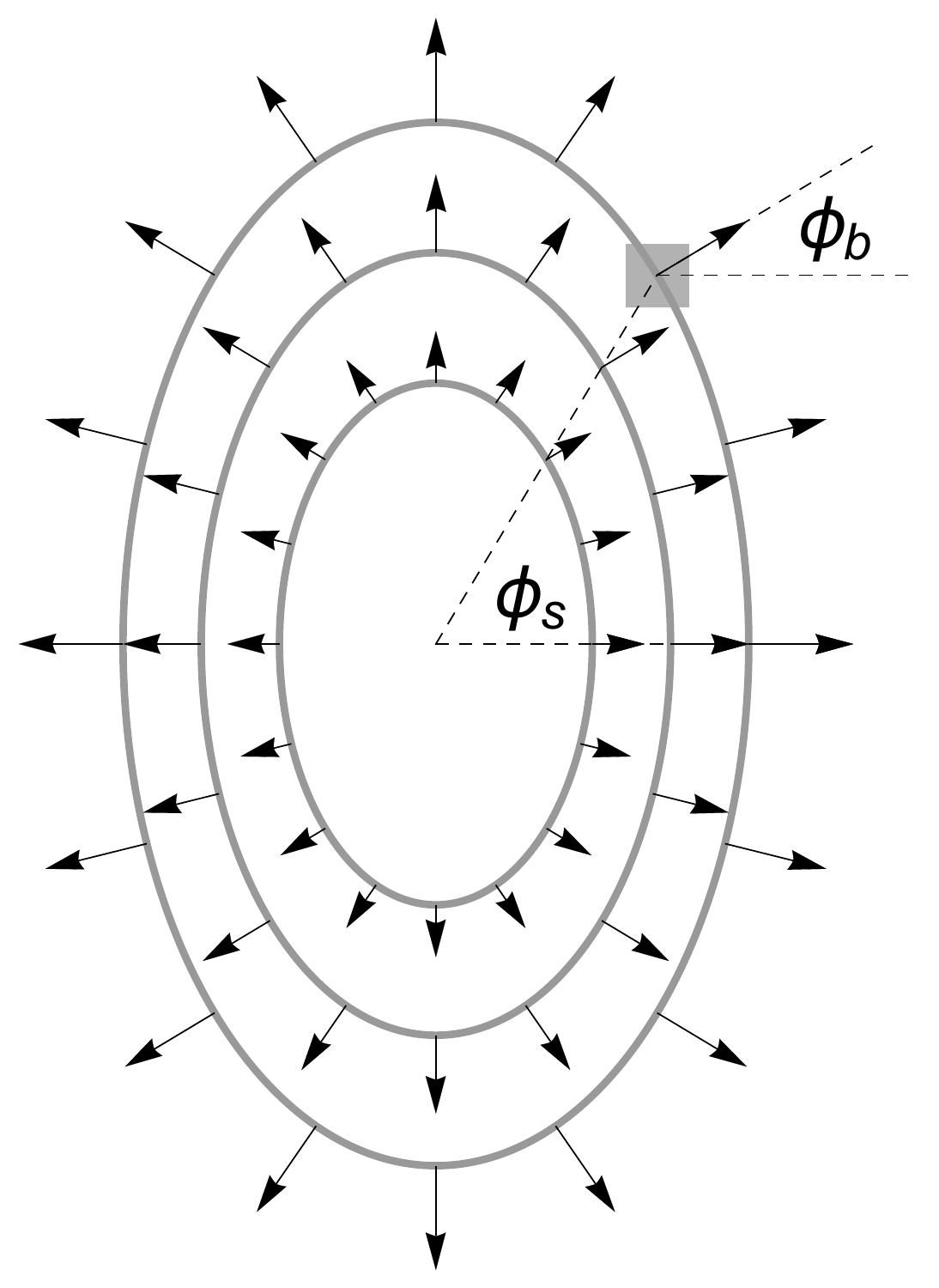}
\caption{\label{fig:sketch_boost_angle} 
Illustration of the elliptical freeze-out surface in the transverse plane assumed in this paper. The angle $\phis$ is the azimuthal angle of the position vector pointing to a source element relative to the reaction plane. The angle $\phib$ describes the direction of the velocity vector of the source element. We assume that $\phib$ is perpendicular to the freeze-out ellipse.}
\end{figure}

Our model closely follows the blast-wave model reported in Refs.~\cite{Schnedermann:1992ra,Schnedermann:1993ws}. Instead of an azimuthally symmetric freeze-out hyper-surface, however, we consider an elliptical shape \cite{Retiere:2003kf}. This is in line with measurements of the freeze-out geometry through azimuthally-differential pion femtoscopy \cite{Adams:2003ra,Adamova:2017opl} and hydrodynamic 
models. In our model, the flow velocity vector of a fluid cell points in the direction perpendicular to the freeze-out ellipse. From this, we derive formulas for $\pt$ spectra and $v_2(p_{T})$ of hadron yields that follow from the Cooper-Frye freeze-out formula \cite{Cooper:1974mv} without further assumptions.  

Assuming a certain collective flow velocity at the space-time coordinates at which fluid cells freeze out and form hadrons, the blast wave model describes the final stage of the hydrodynamic evolution. The produced hadrons are then assumed to freely stream to the detector without further interaction. Often, the contribution of resonance decays to particle spectra is neglected. The fit range is then chosen carefully to minimize the effect of feed-down on the 
extracted fit parameters. Several authors also included resonance decays in the blast-wave 
fits \cite{Broniowski:2001we,Melo:2019mpn,Mazeliauskas:2019ifr}. Recently, resonance decays were even taken into account in a fit of a state-of-the-art hydrodynamic model to transverse momentum spectra measured at the LHC \cite{Devetak:2019lsk}.

Early on it was realized that an azimuthal modulation of the collective radial flow 
velocity provides a natural explanation for the experimentally observed azimuthal anisotropies of particle yields relative to the reaction plane \cite{Huovinen:2001cy}. In Ref.~\cite{Huovinen:2001cy}, a compact formula for the flow coefficient $v_2(\pt)$ was derived. In contrast to the blast-wave formula for the transverse momentum spectra in Ref.~\cite{Schnedermann:1993ws}, a radial dependence of the collective flow velocity was not considered.

The STAR Collaboration used a generalized version of the $v_2(\pt)$ formula (``STAR formula'') of Ref.~\cite{Huovinen:2001cy} to fit $v_2$ data for identified hadrons \cite{Adler:2001nb, Adamczyk:2015fum}. The STAR formula provides a good description of the transverse momentum dependence of $v_2$ for various mesons and baryons for center-of-mass energies ranging from 7.7~GeV up to LHC energies \cite{Sun:2014rda,Acharya:2017dmc}. Moreover, in Ref.~\cite{Acharya:2017dmc} it was shown that the $v_2(\pt)$ of deuterons is nicely described with parameters obtained from a simultaneous fit to spectra and $v_2(\pt)$ for pions, kaons, and protons. The STAR formula provides a better fit to the data than the formula of Ref.~\cite{Huovinen:2001cy}. The difference with respect to the formula in Ref.~\cite{Huovinen:2001cy} is a factor $[1 + 2 s_2 \cos(2 \phi)]$ which describes the azimuthal density of source elements. This is an additional assumption that does not directly follow from the Cooper-Frye ansatz. 

An expression for $v_2(\pt)$ taking also the radial dependence of the flow velocity into account was reported in Ref.~\cite{Song:2010er}. In this formula, the direction of the boost velocity of a fluid cell is assumed to be parallel to the position vector pointing to the source element. In Ref.~\cite{Retiere:2003kf}, it was pointed out that a  natural generalization of the blast-wave model for noncentral collisions is to assume that the boost vector is perpendicular to the elliptical subshell on which the source element is found. This is illustrated in Fig.~\ref{fig:sketch_boost_angle}. Using this assumption, we derive a compact formula for $v_2(\pt)$ in this paper. 

Following Refs.~\cite{Ruuskanen:1986py,Schnedermann:1993ws}, we assume that the velocity of a source element in the laboratory system is given by 
\begin{align}
    u^\mu = \cosh \rho \, (\cosh \eta, \tanh \rho \cos \phib, \tanh \rho \sin \phib, \sinh \eta).
    \label{eq:boost_velocity}
\end{align}
Here, $\eta$ is the space-time rapidity in the longitudinal direction and $\rho$ is the transverse rapidity ($\rho = \arctanh \betat$) where $\betat$ is the transverse velocity).

For the integration over the elliptical freeze-out surface, we introduce the variables $\rhat$ and $\phihat$ so that the position of a source element in the transverse plane is given by
\begin{align}
    x = R_x \rhat \cos \phihat, \quad
    y = R_y \rhat \sin \phihat,
    \label{eq:rhat_phihat}
\end{align}
where $R_x$ and $R_y$ are the radii of the ellipse along the $x$ and $y$ axes, respectively. An area element on the ellipse is then given by $\rmd A_\mathrm{T} = R_x R_y \rhat \rmd \rhat \rmd \phihat$. In our model, the angles $\phib$, $\phis$, and $\phihat$ [cf.~Fig.~\ref{fig:sketch_boost_angle} and Eq.~\ref{eq:rhat_phihat}] are related by
\begin{equation}
    \tan \phib = \frac{R_x^2}{R_y^2} \tan \phis = \frac{R_x}{R_y} \tan \phihat.
    \label{eq:phib_phis_phihat}
\end{equation}

The freeze-out time in the laboratory system $t_\mathrm{f} = \sqrt{\tau_\mathrm{f} + z^2}$ is defined by a fixed longitudinal proper time $\tau_\mathrm{f}$. In each longitudinally moving system, freeze-out happens instantaneously, independent of the radial coordinate, i.e., $\rmd \tau_\mathrm{f}(r) / \rmd r \equiv 0$. This is a frequently made assumption when the blast-wave model is used to fit experimental data \cite{Abelev:2008ab,Abelev:2013vea,Khachatryan:2016yru,alice_pi_k_p_spectra_5tev}. The effect of using a different freeze-out surface $\tau_\mathrm{f} = \tau_\mathrm{f}(\rhat)$ is discussed in Appendix~\ref{sec:freeze_out_surface}. Assuming an elliptical freeze-out hyper-surface, the invariant particle yield averaged over the azimuthal angle $\phip$ of the particle is given by
\begin{align}
    \frac{1}{2 \pi p_\mathrm{T}} \frac{\rmd N}{\rmd p_\mathrm{T} \rmd y}
    & \propto
    \mt \int \limits_{0}^{1} \rhat \rmd \rhat  \int \limits_{0}^{2 \pi} \rmd \phihat  \;
    I_0(\xi_p) K_1(\xi_m), \label{eq:inv_yield} \\
    \xip & \equiv \xip(\rhat, \phihat) = \frac{\pt \sinh \rho(\rhat, \phihat)}{T},\\
    \xim & \equiv \xim(\rhat, \phihat) = \frac{\mt \cosh \rho(\rhat, \phihat)}{T},
\end{align}
with $I_0$ and $K_1$ being modified Bessel functions. The radial velocity profile is taken to be
\begin{align}
    \rho &\equiv \rho(\rhat, \phihat) = \rhat \left(\rho_0 + \rho_2 \cos(2 \phib) \right) 
    \label{eq:rho}\\
    \phib & \equiv \phib(\phihat) = \arctan\left(\frac{R_x}{R_y} \tan \phihat \right) 
    + \lfloor \frac{\phihat}{\pi} + \frac{1}{2} \rfloor \pi \label{eq:phib}
\end{align}
where $\lfloor x \rfloor$ denotes the greatest integer less than or equal to $x$.
The corresponding expression for $v_2(\pt)$ is obtained as the average of $\cos(2 \phip)$ over the azimuthal particle distribution $\rmd N/\rmd \phip$. The integrations over $\phip$ and the space-time rapidity $\eta$ can be done analytically, resulting in 
\begin{align}
    v_2(\pt) = 
    \frac{\int_{0}^{1} \rhat \rmd \rhat \int_{0}^{2 \pi} \rmd \phihat  \; 
    I_2(\xip) K_1(\xim) \cos (2 \phib)}
    {\int_{0}^{1} \rhat \rmd \rhat \int_{0}^{2 \pi} \rmd \phihat  \; 
    I _0(\xip) K_1(\xim)}.
    \label{eq:v2}
\end{align}
The integration over $\rhat$ and $\phihat$ is then performed numerically \cite{github_bw}. The functions $I_0$, $I_2$ in Eqs.~(\ref{eq:inv_yield}) and (\ref{eq:v2}) are modified Bessel functions of the first kind and $K_1$ is a modified Bessel function of the second kind. Equation~(\ref{eq:v2}) differs from the formula in Ref.~\cite{Song:2010er} as the boost direction of a source element in our model is a function of $\phihat$ (or $\phis$) and, in general, does not coincide with the direction of the position vector pointing to the source element ($\phib \neq \phis$). More details on the derivation of these equations are given in Appendix~\ref{sec:cooper-frye-ansatz}. 


\section{\label{sec:feed_down} Feed-down from resonance decays}

\begin{figure}[htbp] 
\begin{overpic}[bb = 0 0 600 820,clip,width=0.495\textwidth]{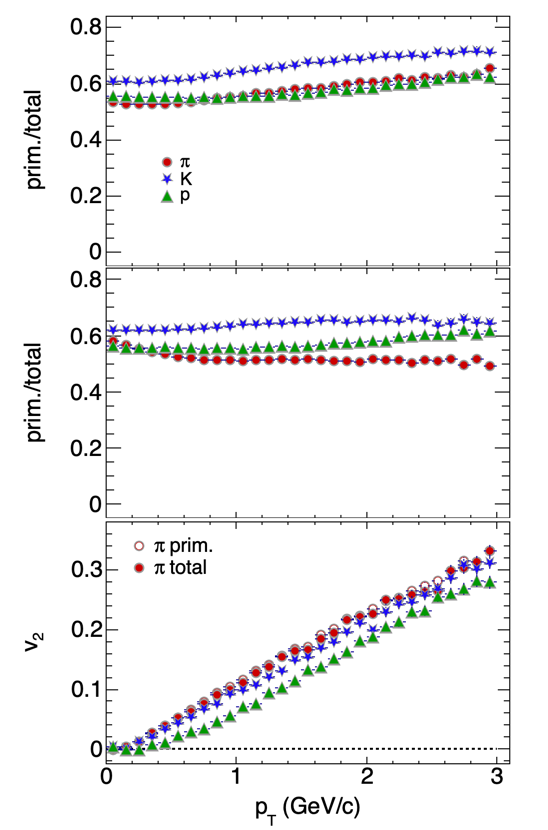} 
\put(15,85){\Large \bf {a)}}
\put(15,45){\Large \bf {b)}}
\put(15,24){\Large \bf {c)}}
\end{overpic}
\caption{(Color online) Feed-down contributions to the yield (a) for the blast-wave parameters $T = \unit[128]{MeV}$, $\rho_{0}=1.05$, $\rho_{2}=0.09$, $R_{x}/R_{y}=0.83$ and (b) for $T = \unit[109]{MeV}$, $\rho_{0}=1.11$, $\rho_{2}=0.077$, $R_{x}/R_{y}=0.83$ for for $\pi$, K, and p.
Panel (c) shows the elliptic flow for primary plus secondary particles (total) as filled symbols for the parameters of panel (a). In addition, the primary $v_{2}$ for pions is shown as open symbols.}
\label{fig:feed_down}
\end{figure}

The blast-wave fits in this paper only take into account primary hadrons, i.e., hadrons directly produced from the thermal source. The experimental measurement, however, includes contributions from the decay of short-lived hadrons. Pions are affected most by feed-down since they are the lightest hadrons and are therefore at the end of any decay chain. Examples for feed-down contribution to the pions are the decays $\eta \to \pi^+\pi^-\pi^0$, $\rho \to \pi \pi$, and $\Delta \to N\pi$. Feed-down contributions affect the shape of the measured transverse momentum spectra and the measured elliptic flow $v_2(\pt)$. This makes the interpretation of the fit parameters of the blast-wave fit with only primary particles potentially difficult. We therefore estimate a possible change of the transverse momentum spectra and $v_2(\pt)$ due to feed-down by means of a Monte Carlo study.

In our feed-down Monte Carlo simulation, we generate primary hadrons in the rest frame of a source element isotropically with momenta following a Boltzmann distribution for a kinetic freeze-out temperature $T$. The relative abundances of different hadron species are determined by the particle number densities in a noninteracting hadron gas for a vanishing chemical potential $\mu$ in the Boltzmann approximation as given by 
\begin{align}
n \propto g m^2 T_\mathrm{ch} K_2\left(\frac{m}{T_\mathrm{ch}}\right),
\end{align}
where $g$ denotes the spin degeneracy and $m$ indicates the particle mass. $K_2$ is a modified Bessel function of the second kind. We use a chemical freeze-out temperature of $T_\mathrm{ch} = \unit[156]{MeV}$~\cite{Andronic:2017pug}.
All hadrons implemented in the \Py~8 event generator~\cite{Sjostrand:2006za,Sjostrand:2007gs} up to a mass of 
$\unit[2]{GeV}/c^2$ are considered. The velocity of a source element is described by Eq.~(\ref{eq:boost_velocity}). 

We uniformly distribute points described by $\rhat$ and $\phihat$ over the freeze-out ellipse
and determine the transverse rapidity $\rho$ and the boost direction $\phib$ according to Eqs.~(\ref{eq:rho)} and (\ref{eq:phib}), respectively. The longitudinal rapidity of a source element is drawn from a uniform distribution. The primary hadrons generated in the rest frame of the source element are boosted to the laboratory frame. We then use \Py~8 to simulate the hadron decays where particles with a mean proper decay length above $c \tau = \unit[1]{mm}$ are considered stable.   
\begin{figure*}[htbp] 
\begin{overpic}[bb = 0 0 600 450,clip,width=0.485\textwidth]{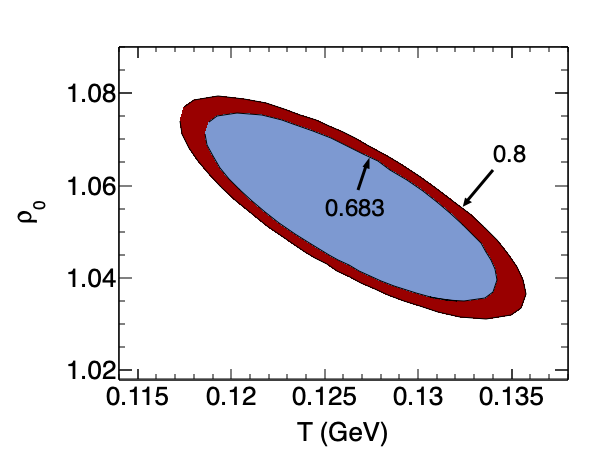} 
\put(24,64){\Large \bf {a)}}
\end{overpic}
\begin{overpic}[bb = 0 0 600 450,clip,width=0.485\textwidth]{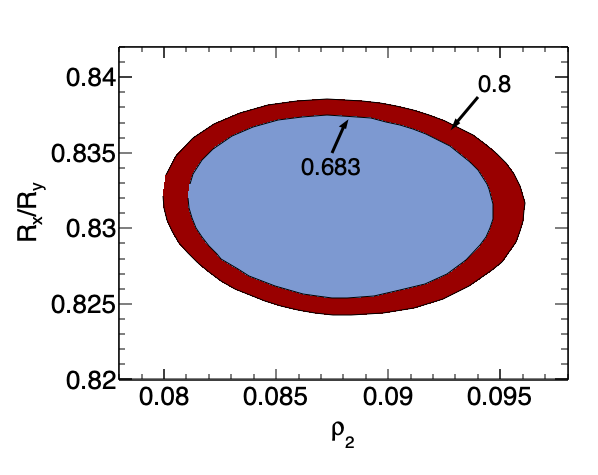} 
\put(24,64){\Large \bf {b)}}
\end{overpic}
\caption{(Color online) Correlation ellipses for (a) $\rho_{0}$ vs. T and (b) $R_{x}/R_{y}$ vs $\rho_{2}$ for the confidence levels 0.683 ($1\sigma$ contour) and 0.8.}
\label{fig:ellipses}
\end{figure*}

Figure~\ref{fig:feed_down} depicts the results of the feed-down calculations for $\pi$, K, and p. The ratio of the primary particle yield to the total yield (primary + secondary) is shown in Fig.~\ref{fig:feed_down}(a) and \ref{fig:feed_down}(b) for two sets of blast-wave parameters (see caption for details). Pions exhibit the largest change in the primary/total ratio between the two parameter sets. In particular, the change at low $\pt$ is larger than for the other particles. The elliptic flow for primary plus secondary particles is shown in Fig.~\ref{fig:feed_down}(c). The primary $v_{2}$ is shown in addition for pions. Overall the difference between the primary and total $v_2(p_T)$ for pions, kaons, and protons is negligible for our analysis. For pions resonance contributions are expected to give rise a small reduction of the total pion $v_2(\pt)$ at low $\pt$ ($\lesssim 0.5\,\mathrm{GeV}/c$) \cite{Hirano:2002ds}. The reduction of $v_2(\pt)$ in our model is found to be compatible with results of other calculations~\cite{Dong:2004ve,Nasim:2014iea}. 

We conclude that for $v_{2}(\pt)$ no special feed-down correction is needed while for the particle yields, the $\pt$ dependent feed-down contribution is sensitive to the particular values of the blast-wave parameters. For the studied parameters and particles, the variation of the ratio over the $\pt$ range up to 3\,GeV/$c$ is less than $\pm 8\%$ which is not much larger than the total data uncertainties at low $\pt$. Considering the experimental uncertainties and the limitations of the blast-wave approach, we decided to not take those model based feed-down effects into account for the current study. 


\section{\label{sec:results} Results}

To reduce the impact of feed-down, we therefore decided, based on the discussion in Section~\ref{sec:feed_down}, to exclude the pions entirely from the fits. In addition, we varied the fit ranges to test the stability of the results. With increasing transverse momentum, jet contributions are getting more relevant. Jets are produced in a heavy-ion collision during the first hard scatterings of initial partons. The resulting high-$\pt$ hadrons participate only little in the bulk evolution of the matter. Their kinematic distributions are therefore not covered by the blast-wave model. In order to minimize the influence of jet contributions, we limited the $\pt$ ranges for the fit to a value $\pt^\mathrm{max}$:  
\begin{equation}
\pt^\mathrm{max} = m c \gamma^\mathrm{max}\beta^\mathrm{max} + 1.0\, \mathrm{GeV}/c   
\label{eqn:ptmax}
\end{equation}
where $m$ is the rest mass of the particles and $\beta^\mathrm{max}$ is set to 0.68. There is some arbitrariness in the choice of the fit range. The change in transverse momentum due to radial flow increases with the particle's mass and hence jet contributions are expected to become relevant at a higher $p_T$ for particles with larger masses. This is taken into account by our choice of the fit range. Moreover, we ensure that the maximum of the $\rmd^2N/\rmd p_T \rmd y$ distribution lies within the fit range.

\begin{table*}[t] 
\centering
\caption{Used elliptic flow ($v_{2}$) and $\rmd N/\rmd\pt$ data. The upper fit range is according to Eq.~\ref{eqn:ptmax}.
\label{tab:data}}
\begin{tabular}{ l c c c c c c c c c}		
   & $\boldsymbol{\pi}$ & $\boldsymbol{K}$ & $\boldsymbol{p}$ & $\boldsymbol{\phi}$ & $\boldsymbol{\Omega}$ & $\boldsymbol{D^{0}}$ & $\boldsymbol{d}$ & $\boldsymbol{J/\psi}$ & $\boldsymbol{\Upsilon}$ \\
  \hline
  $\boldsymbol{v_{2}}$ reference & \cite{Abelev:2014pua} & \cite{Abelev:2014pua} & \cite{Abelev:2014pua} & \cite{Abelev:2014pua} & \cite{Abelev:2014pua} & \cite{Abelev:2013lca} & \cite{Acharya:2017dmc} & \cite{ALICE:2013xna} & \cite{Acharya:2019hlv} \\
  $\sqrt{s_\mathrm{NN}}$ (TeV) & 2.76 & 2.76 & 2.76 & 2.76 & 2.76 & 2.76 & 2.76 & 2.76 & 5.02 \\
  Centrality (\%) & 30-40 & 30-40 & 30-40 & 30-40 & 30-40 & 30-50 & 30-40 & 20-40 & 5-60 \\
  Acceptance & $|y|<0.5$ & $|y|<0.5$ & $|y|<0.5$ & $|y|<0.5$ & $|y|<0.5$ & $|y|<0.5$ & $|y|<0.5$ & $2.5<y<4$ & $2.5<y<4$ \\ \hline
  
   $\boldsymbol{\rmd N/\rmd \pt}$ reference & \cite{Adam:2015kca} & \cite{Adam:2015kca} & \cite{Adam:2015kca} & \cite{Adam:2017zbf} & \cite{ABELEV:2013zaa} & \cite{Adam:2015sza} & \cite{Adam:2015vda,Acharya:2017dmc} & & \cite{Khachatryan:2016xxp} \\
  $\sqrt{s_\mathrm{NN}}$ (TeV) & 2.76 & 2.76 & 2.76 & 2.76 & 2.76 & 2.76 & 2.76 &  & 2.76 \\
  Centrality (\%) & 30-40 & 30-40 & 30-40 & 30-40 & 20-40 & 30-50 & 20-40 &  & 0-100 \\
  Acceptance & $|\eta|<0.8$ & $|\eta|<0.8$ & $|\eta|<0.8$ & $|y|<0.5$ & $|y|<0.5$ & $|y|<0.5$ &$|y|<0.5$ &  & $|y|<2.4$ \\ \hline
  
  $\pt$ fit range (GeV/$c$) & [0.4,1.1] & [0.2,1.5] & [0.2,1.9] & [0.5,1.9] & [1.1,2.6] & [0.9,2.7] & [0.5,2.7] & [0.2,3.9] & [0.1,9.8]
\end{tabular}
\end{table*}

\begin{table*}[t] 
\centering
\caption{Blast-wave fit results.
\label{tab:results}}
\begin{tabular}{ p{2cm} p{2.5cm} p{2cm} p{2cm} p{2.3cm} p{2cm} }	
\bf{PID} & \bf{Comment} & \bf{$\boldsymbol{T}$ (MeV)} & $\boldsymbol{\rho_{0}}$ & $\boldsymbol{\rho_{2}}$ & $\boldsymbol{R_{x}/R_{y}}$ \\ \hline
\multicolumn{6}{c}{ALICE fit range~\cite{Abelev:2013vea}} \\
$\pi$,K,p & $\rmd N/\rmd \pt$ only & $107.5\pm2.2$ & $1.13\pm0.03$ & $0.41\pm0.03$ & $0.43\pm0.04$ \\
$\pi$,K,p & $\rmd N/\rmd \pt$ + $v_{2}$ & $107.5\pm2.2$ & $1.13\pm0.01$ & $0.076\pm0.002$ & $0.83\pm0.003$ \\ \hline
\multicolumn{6}{c}{Fit range based on Eqn.~\ref{eqn:ptmax} - standard fit} \\
K,p,$\phi$,$\Omega$ & $\beta^\mathrm{max}=0.68$ & $128.4\pm5.1$ & $1.05\pm0.01$ & $0.09\pm0.004$ & $0.83\pm0.004$ \\ \hline
\multicolumn{6}{c}{Fit range based on Eqn.~\ref{eqn:ptmax} - variations} \\
K,p,$\phi$,$\Omega$ & $\beta^\mathrm{max}=0.75$ & $131.2\pm4.7$ & $1.05\pm0.01$ & $0.093\pm0.004$ & $0.83\pm0.004$ \\
K,p,$\phi$,$\Omega$ & $\beta^\mathrm{max}=0.55$ & $121.0\pm6.5$ & $1.07\pm0.02$ & $0.089\pm0.006$ & $0.84\pm0.004$ \\
K,p,$\phi$,$\Omega$,d & $\beta^\mathrm{max}=0.68$ & $134.1\pm3.1$ & $1.04\pm0.01$ & $0.093\pm0.004$ & $0.83\pm0.003$ \\
K,p,$\phi$,$\Omega$,$D^{0}$ & $\beta^\mathrm{max}=0.68$ & $128.7\pm5.2$ & $1.05\pm0.01$ & $0.09\pm0.005$ & $0.83\pm0.004$ \\
K,p,$\phi$,$\Omega$,$\pi$ & $\beta^\mathrm{max}=0.68$ & $109.3\pm2.2$ & $1.11\pm0.01$ & $0.077\pm0.002$ & $0.83\pm0.003$ \\
\end{tabular}
\end{table*}

\begin{figure*}[htbp] 
\begin{overpic}[bb = 50 50 1400 550,clip,width=0.9\textwidth]{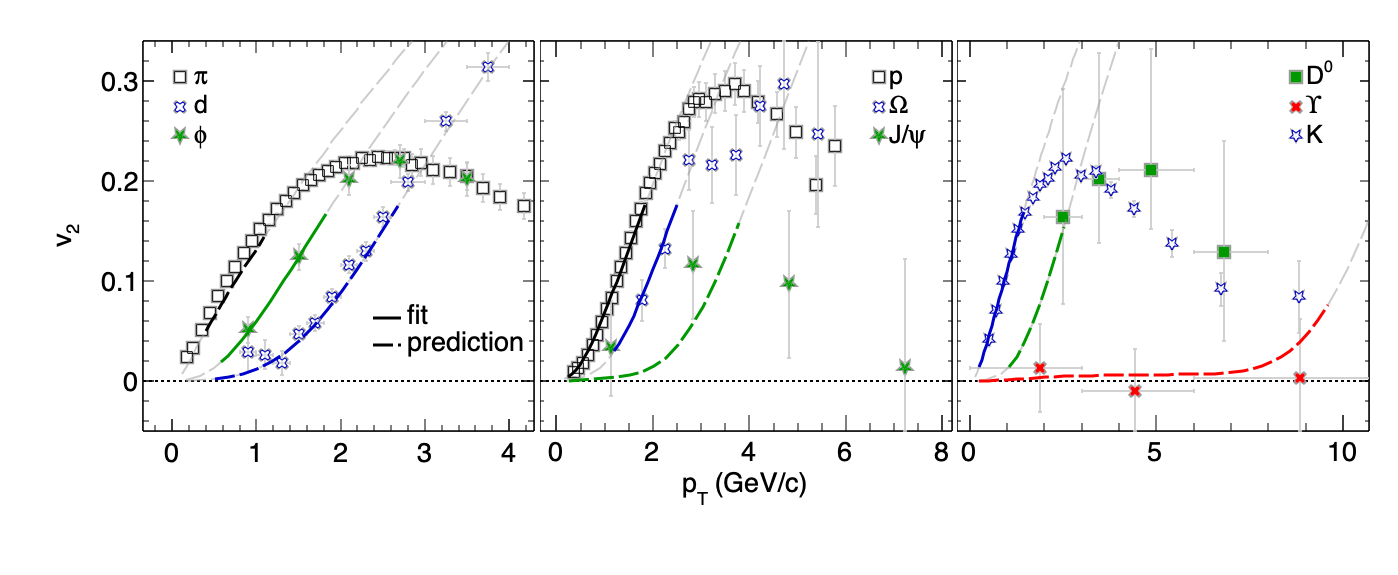} 
\end{overpic}

\vspace{1cm}

\begin{overpic}[bb = -100 0 1000 
850,clip,width=0.9\textwidth]
{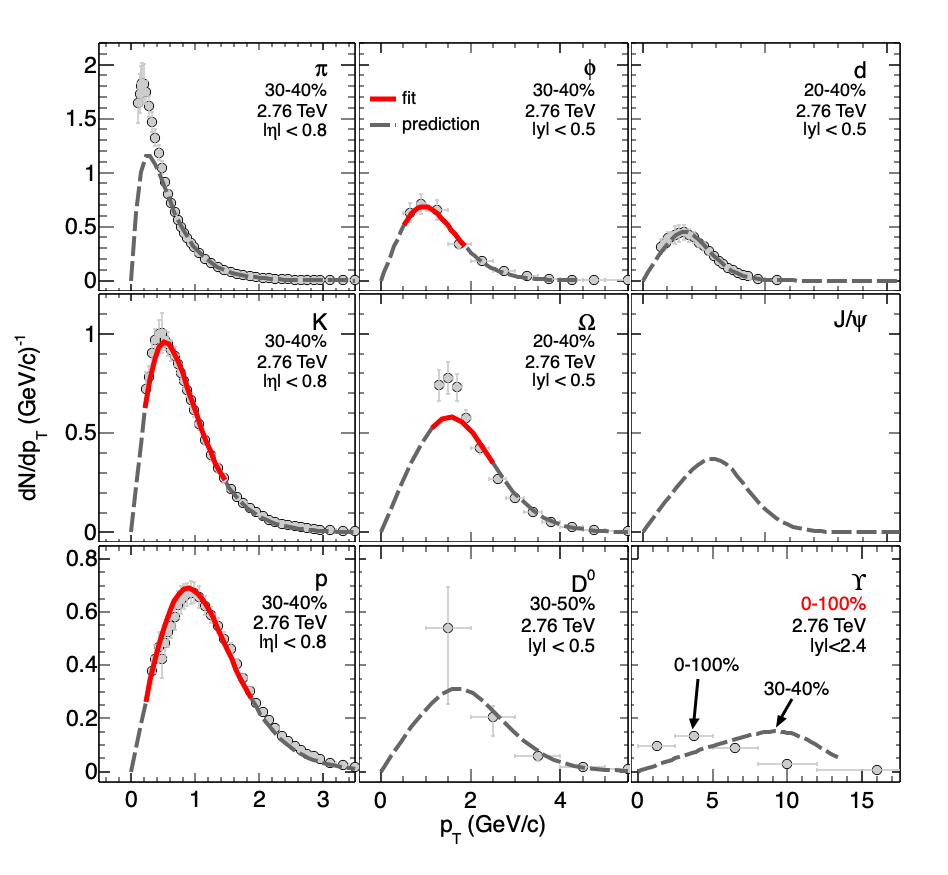} 
\end{overpic}
\caption{(Color online) Upper: Result of the simultaneous blast-wave fit and corresponding predictions for identified particle $v_{2}$ data. Gray lines represent extrapolations outside the fitted $\pt$ range. Lower: Corresponding results for the normalized identified particle spectra. Only the $\pt$ ranges depicted by red-solid lines are used for the simultaneous fit. The used centrality ranges for $v_{2}$ are listed in Tab.~\ref{tab:data}.}
\label{fig:v2_fits}
\end{figure*}

Because of the limited availability of data, a certain mix of centralities, collision energies, and detector acceptances for the identified particles had to be used. In Table~\ref{tab:data}, an overview of the used data is given. The slightly different centralities for some particle species in ranges 30--40\% ($\pi$, $p$, $K$, $\phi$), 30--50\% ($D^{0}$, $J/\psi$) and 20--40\% ($d$ and $\Omega$) have little impact on the overall results but might be responsible for smaller deviations. For the $\Upsilon$ $\rmd N/\rmd \pt$ spectrum, which is only available for minimum bias (0--100\%) collisions, large deviations are expected and therefore the spectrum is only listed and shown for reference. The peak maximum is at about 4\,GeV/$c$ in $\pt$ which is, as expected, compatible with a transverse flow rapidity of about 0.5 in peripheral collisions. For the particles used in the fit, the collision system and energy is Pb--Pb at $\sqrt{s_{NN}} =  \unit[2.76]{TeV}$, while for $\Upsilon$  $\sqrt{s_{NN}} = \unit[5.02]{TeV}$ data were used for $v_{2}$. For the $\Upsilon$, the used elliptic flow data are expected to be an upper limit with respect to $\sqrt{s_{NN}} = \unit[2.76]{TeV}$. Due to the general assumption of a boost-invariant scenario, no large differences are expected from the different rapidity acceptances as listed in Table~\ref{tab:data}. For the data points, statistical and systematic uncertainties were added in quadrature. All particle spectra were normalized to their integral within the data range.

To test some of the assumptions made, we varied the mix of particles used in the fit and their $\pt$ fit ranges. The results are listed in Table~\ref{tab:results}. For most of the variations, the fit parameters changed only in the order of the fit error bars. The biggest impact on the temperature and the radial flow rapidity occurs when the pions are included, even if their fit range was limited to the narrow window of $0.4 < \pt < \unit[1.1]{GeV}/c$. As discussed in Section~\ref{sec:feed_down}, the pion spectra are affected most by feed-down contributions, while their statistical power is dominating the simultaneous fit. For that reason, we excluded them from the fit.

Figure~\ref{fig:ellipses} depicts the correlation ellipses for $T$ versus $\rho_{0}$ and $\rho_{2}$ versus $R_{x}/R_{y}$ for the standard fit as shown in Fig.~\ref{fig:v2_fits}. A clear anti-correlation between the temperature and the radial expansion rapidity is visible, while $\rho_{s}$ and $R_{x}/R_{y}$ do not show a significant correlation. The latter is somehow unexpected since both parameters generate $v_{2}$. A comparison to the blast-wave results from~\cite{Abelev:2013vea} with a different fit function, using only $\pi$, K, and p $\rmd N/\rmd \pt$ data, is also listed in Tab.~\ref{tab:results}. The temperature is with 107.5\,MeV almost identical to that with 106\,MeV from~\cite{Abelev:2013vea} while the transverse surface velocity $\beta_{s}$ deviates by about 10\%. 

For the standard fit, the line shapes for particle yields and elliptic flow data are depicted in Fig.~\ref{fig:v2_fits}. Solid lines indicate the particles and $\pt$ ranges used for the simultaneous fit. Dashed lines are predictions by the model with the particles mass being the only free parameter left. For the reasons pointed out in Section~\ref{sec:results}, a perfect description of the data cannot be expected. Nonetheless, most of the fitted data points for $v_{2}(\pt)$ have a less than $1\sigma$ deviation from the fit. For the $\rmd N/\rmd \pt$ spectra, the deviations are slightly bigger, but those are expected due to the feed-down contributions. A similar situation appears for the predicted distributions. In the $\pt$ range lower than the drop in $v_{2}$, which is most likely due to jet contributions, the data points in $v_{2}(\pt)$ and $\rmd N/\rmd \pt$ are mostly within $1\sigma$ of the prediction. The exception are the pions which show a constant offset by about 100\,MeV in $\pt$ or 0.015 in $v_{2}$, respectively. Including the pions in the fit reduces the pion $\chi^{2}/$point in $v_{2}$ from 18.5 to 3.8 but at the same time the description of the other particles gets slightly worse. Advanced feed-down calculations are needed to study those effects in detail. 

The prediction for the $\Upsilon$ and its perfect agreement with the recent ALICE data is remarkable because it shows that no deviation from zero in $v_{2}$ is expected in the blast-wave model within the current kinematic reach of the data. At higher transverse momenta above 9\,GeV/$c$, an increasing elliptic flow is expected, following the systematics of the mass ordering of $v_2(\pt)$ observed for lighter particles \cite{Adams:2004bi,Ollitrault:2008zz}. The late rise in $p_{T}$ of the $\Upsilon$ $v_{2}$ is mainly a kinematic effect due to its large mass of 9.46\,GeV.


\section{\label{sec:summary} Summary and conclusions}
A simultaneous blast-wave fit to elliptic flow data and particle yields at LHC energies for various particles species was presented. A compact formula for $v_2(\pt)$ for an elliptical freeze-out hyper-surface was derived in this paper and used in the fit. Studies on feed-down from resonance decays showed that their effect on $v_2(\pt)$ is small. Moreover, the effect of variations of the $\pt$ fit ranges and the composition of particles included in the fit on the fit parameters were studies. Stable fit results were achieved by excluding the pions and using only K, p, $\phi$ and $\Omega$. 

The predicted elliptic flow distributions for heavy particles like the $D^{0}$, $J/\psi$, and $\Upsilon$ are in good agreement with the recent data from the ALICE Collaboration. Especially the result of the $\Upsilon$ $v_{2}$ shows that within the current kinematic reach no significant deviation from zero is expected in the blast-wave model. 

To address the question of to what extent also $b$ quarks thermalize one needs to consider both the $\pt$ spectra and the elliptic flow of $\Upsilon$ mesons. $\Upsilon$ $\pt$ spectra are currently available only for the 0--100\% centrality class and it is hard to judge whether these spectra agree with a blast-wave model prediction or not. With the current $\Upsilon$ $v_{2}$ data alone one cannot rule out a scenario in which also $b$ quarks thermalize and the $\Upsilon$ mesons follow the $v_2(\pt)$ mass ordering of lighter particles resulting from a collective expansion of the medium.  

\begin{acknowledgments}
We thank Peter Braun-Munzinger, Markus K\"{o}hler, Yvonne Pachmayer, Kai Schweda, Johanna Stachel, and Martin V\"{o}lkl for valuable discussions. This work is part of and supported by the DFG Collaborative Research Centre ``SFB 1225
(ISOQUANT)''.
\end{acknowledgments}

\appendix

\section{Cooper-Frye ansatz for an elliptical freeze-out hyper-surface}
\label{sec:cooper-frye-ansatz}
Equations (\ref{eq:inv_yield}) and (\ref{eq:v2}) used for simultaneously fitting particle spectra and $v_2$ follow from the Cooper-Frye freeze-out formula \cite{Cooper:1974mv}. In our model the freeze-out hyper-surface is an elliptical cylinder along the beam direction which can be parameterized by
\begin{align}
    \Sigma_\mathrm{f}^\mu(\rhat, \phihat) = (t_\mathrm{f}(\rhat, z), \rhat  R_x \cos \phihat, \rhat R_y \sin \phihat, z)
\end{align}
where $R_x$ and $R_y$ are the radii of the freeze-out ellipse in the transverse plane in the $x$ and $y$ directions, respectively, and $t_\mathrm{f}(\rhat, z)$ is the freeze-out time in the laboratory system. 
As outlined, e.g., in Refs.~\cite{Schnedermann:1992ra,Schnedermann:1993ws,Yagi:2005yb,Florkowski:2010zz}
the normal vector to the surface is then given by
\begin{align}
    \rmd \Sigma_\mu & = \epsilon_{\mu \nu \lambda \rho} 
    \frac{\partial \Sigma^\nu}{\partial \hat r} 
    \frac{\partial \Sigma^\lambda}{\partial \hat \phi}
    \frac{\partial \Sigma^\rho}{\partial z}
    \,  \rmd \rhat \, \rmd \phihat \, \rmd z \\ \nonumber
    & = \left(
    1, 
    -\frac{1}{R_x} \frac{\partial \Sigma^0}{\partial \hat r} \cos \hat \phi,
    -\frac{1}{R_y} \frac{\partial \Sigma^0}{\partial \hat r} \sin \hat \phi,
    - \frac{\partial \Sigma^0}{\partial z} 
    \right) \\
    & \quad \;  R_x R_y \rhat  \rmd \rhat \, \rmd \phihat \, \rmd z \nonumber.
    \label{eq:normal_vector}
 \end{align}
The freeze-out time in the laboratory system is taken to be $t_\mathrm{f} = \sqrt{\tau_\mathrm{f}(\rhat)^2 + z^2}$
where $\tau_\mathrm{f}(\rhat)$ is the freeze-out time in a system moving along the 
beam direction. Moreover, we assume instantaneous radial freeze-out, i.e., 
$\tau_\mathrm{f}(\rhat) \equiv \tau_\mathrm{f}$. Using $z = \tau_\mathrm{f} \sinh \eta$, 
and $t_\mathrm{f} = \tau_\mathrm{f} \cosh \eta$, where $\eta$ is the space-time rapidity, 
one obtains $\partial \Sigma^0/\partial z = \tanh \eta$ and Eq.~(\ref{eq:normal_vector})
simplifies to 
\begin{align}
    \rmd\Sigma_\mu 
    = \left(\cosh \eta, 0, 0, \sinh \eta 
 \right) \tau_\mathrm{f} R_x R_y \rhat \rmd \rhat \, \rmd \phihat \, \rmd \eta .
\end{align}
With 
\begin{align}
    p^\mu = (\mt \cosh y, \pt \cos \phip, \pt \sin \phip, \mt \sinh y)
\end{align}
one then obtains
\begin{align}
    p^\mu \,\rmd \Sigma_\mu &= \tau_\mathrm{f} m_T \left(\cosh \eta \cosh y - \sinh y \sinh \eta \right) \\ 
    & \quad \; R_x R_y \rhat \, \rmd \rhat  \, \rmd \phihat \, \rmd \eta . \nonumber
\end{align}
With Eq.~(\ref{eq:boost_velocity}), the product $p^\mu u_\mu$ used in the Cooper-Frye formula 
takes the form
\begin{align}
    p^\mu u_\mu = \mt \cosh(y-\eta) \cosh \rho - \pt \cos(\phib - \phip) \sinh \rho
\end{align}
where the boost velocity $\phib$ of the source element is given by Eq.~(\ref{eq:phib}).
Analytic integration over the azimuthal angle $\phip$ of the particle and the space-time rapidity $\eta$
then yields Eqs.~(\ref{eq:inv_yield}) and (\ref{eq:v2}).

\section{Variation of the freeze-out hyper-surface}
\label{sec:freeze_out_surface}
\begin{figure}
\includegraphics[width = 0.9\linewidth]{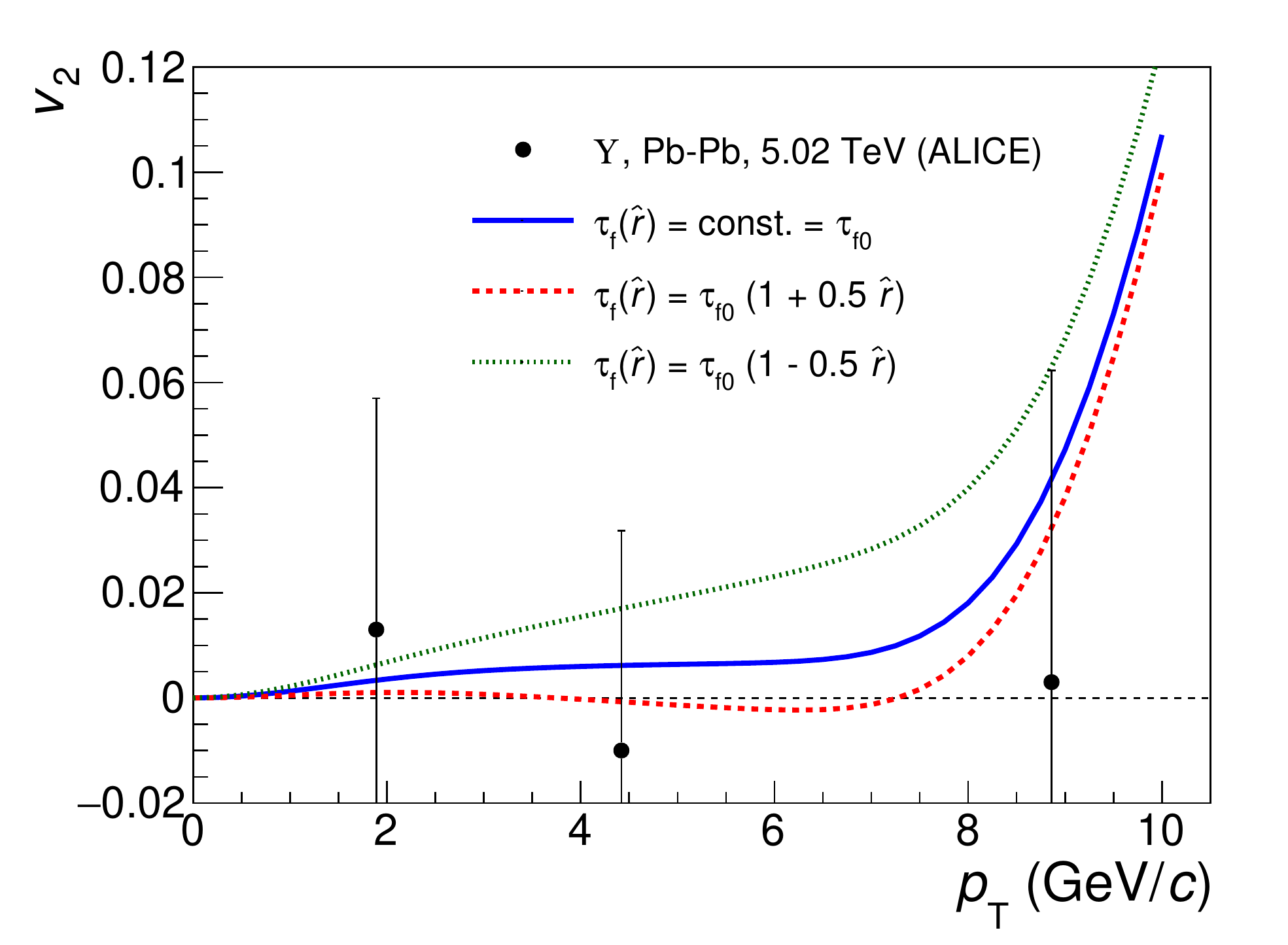}
\caption{Elliptic flow $v_2$ for different freeze-out surfaces for the blast-wave parameters of the standard fit in Table~\ref{tab:results}. In all three cases the model calculation agrees with the ALICE measurement.}
\label{fig:v2_upsilon_different_freeze_out_surfaces}
\end{figure}

In this paper and in many applications of the blast-wave model, an instantaneous freeze-out in the radial direction is assumed. Here we study the more general case $\tau_\mathrm{f} = \tau_\mathrm{f}(\rhat)$. Formulas which generalize Eq.~(\ref{eq:inv_yield}) for the invariant yield and Eq.~(\ref{eq:v2}) for the elliptic flow can still be derived in this case, although the expressions become much longer. In Ref.~\cite{Rybczynski:2012ed} blast-wave model predictions for different freeze-out surfaces were compared to transverse momentum spectra and Hanbury Brown-Twiss radii measured at the LHC. Following this approach we consider the following freeze-out surfaces:
\begin{align}
    \tau_\mathrm{f}(\rhat) =  \tau_\mathrm{f0} (1 + a \rhat) 
\end{align}
with $a = 0, 0.5$, and $-0.5$.
The effect of the different freeze-out surfaces is illustrated in Fig.~\ref{fig:v2_upsilon_different_freeze_out_surfaces} for the case of the elliptic flow of the $\Upsilon$ meson. For lighter particles, the change of the freeze-out surfaces was found to have a smaller effect on the elliptic flow. We conclude that the calculations for all three freeze-out surfaces are consistent with the ALICE measurement.

\bibliography{Blast_wave_fits} 

\end{document}